\documentclass{article}%
\usepackage{amsfonts}
\usepackage{amssymb}
\usepackage{amsmath}
\usepackage{graphicx}%
\providecommand{\U}[1]{\protect\rule{.1in}{.1in}}

\newtheorem{theorem}{Theorem}

\begin{document}

\title{Probability Theory with Superposition Events: A Classical Generalization in
the Direction of Quantum Mechanics}
\author{David Ellerman\\University of Ljubljana, Slovenia}
\date{}
\maketitle

\begin{abstract}
\noindent In finite probability theory, events are subsets $S\subseteq U$ of
the outcome set. Subsets can be represented by $1$-dimensional column vectors.
By extending the representation of events to two dimensional matrices, we can
introduce "superposition events." Probabilities are introduced for classical
events, superposition events, and their mixtures by using density matrices.
Then probabilities for experiments or `measurements' of all these events can
be determined in a manner exactly like in quantum mechanics (QM) using density
matrices. Moreover the transformation of the density matrices induced by the
experiments or `measurements' is the L\"{u}ders mixture operation as in QM.
And finally by moving the machinery into the $n$-dimensional vector space over
$%
\mathbb{Z}
_{2}$, different basis sets become different outcome sets. That
`non-commutative' extension of finite probability theory yields the
pedagogical model of quantum mechanics over $%
\mathbb{Z}
_{2}$ that can model many characteristic non-classical results of QM.

\end{abstract}
\tableofcontents

\section{Introduction: Probability Theory with Superposition Events}

The purpose of this paper is to introduce new concepts such as "superposition
events" into finite probability theory. Let $U=\left\{  u_{1},...,u_{n}%
\right\}  $ be the outcome set or sample space of outcomes with the respective
point probabilities of $p=\left(  p_{1},...,p_{n}\right)  $. Classical events
are represented by subsets $S\subseteq U$ with probabilities $\Pr\left(
S\right)  =\sum_{u_{i}\in S}p_{i}$ where the conditional probability of the
event $T$ given the event $S$ is $\Pr\left(  T|S\right)  =\frac{\Pr\left(
S\cap T\right)  }{\Pr\left(  S\right)  }$.

\section{The Density Matrix Representations}

To generalize classical events to superposition events, we need a richer
mathematical representation than just the notion of a subset. The mathematical
information in a `classical' event $S$ (for convenience, always non-empty)
could be represented in a (normalized) column vector $\left\vert
S\right\rangle $ with $i^{th}$ entry being $\sqrt{\frac{p_{i}}{\Pr\left(
S\right)  }}\chi_{S}\left(  u_{i}\right)  $ (where $\chi_{S}:U\rightarrow
\left\{  0,1\right\}  $ is the characteristic or indicator function for $S$,
$\chi_{S}\left(  u_{i}\right)  =1$ if $u_{i}\in S$ and $0$ otherwise). The
same information could be represented in two dimensions by the diagonal
$n\times n$ matrix $\rho\left(  \Delta S\right)  $ with the diagonal entries
$\frac{p_{i}}{\Pr\left(  S\right)  }\chi_{S}\left(  u_{i}\right)  $, i.e.,

\begin{center}
$\rho\left(  \Delta S\right)  _{i}=\frac{p_{i}}{\Pr\left(  S\right)  }\chi
_{S}\left(  u_{i}\right)  $.
\end{center}

\noindent But the richer two-dimensional matrices allows us to define the
\textit{superposition event} $\Sigma S$ associated with $S$ as being
represented by the $n\times n$ matrix $\rho\left(  \Sigma S\right)  $ (writing
the transpose $\left\vert S\right\rangle ^{t}=\left\langle S\right\vert $) by
multiplying the $n\times1$ column vector $\left\vert S\right\rangle $ times
the $1\times n$ transpose $\left\vert S\right\rangle ^{t}=\left\langle
S\right\vert $:

\begin{center}
$\rho\left(  \Sigma S\right)  =\left\vert S\right\rangle \left\langle
S\right\vert $ with the entries $\rho\left(  \Sigma S\right)  _{ik}%
=\sqrt{\frac{p_{i}}{\Pr\left(  S\right)  }\frac{p_{k}}{\Pr\left(  S\right)  }%
}\chi_{S}\left(  u_{i}\right)  \chi_{S}\left(  u_{k}\right)  $.
\end{center}

\noindent Note that singleton events $S=\left\{  u_{i}\right\}  $ have no
distinct elements to superpose and accordingly $\rho\left(  \Delta\left\{
u_{i}\right\}  \right)  =\rho\left(  \Sigma\left\{  u_{i}\right\}  \right)  $.

Both $\rho\left(  \Delta S\right)  $ and $\rho\left(  \Sigma S\right)  $ are
examples of \textit{real density matrices} which can be defined abstractly as
symmetric matrices $\rho=\rho^{t}$ over the reals with trace (sum of diagonal
elements) $\operatorname{tr}\left[  \rho\right]  =1,$ and with non-negative
eigenvalues. But for practical purposes, density matrices (over the reals
unless otherwise stated) may be taken to be any probabilistic mixtures of
matrices of the form $\rho\left(  \Sigma S\right)  $. That is, for any
probability distribution $q=\left(  q_{1},...,q_{m}\right)  $ and classical
events $S_{j}\subseteq U$ for $j=1,...,m$, the convex combination $\sum
_{j=1}^{m}q_{j}\rho\left(  \Sigma S_{j}\right)  $ is also a density matrix.

A density matrix $\rho$ is said to be \textit{pure} if $\rho^{2}=\rho$, and
otherwise \textit{mixed}. For instance, $\rho\left(  \Sigma S\right)  $ is
pure while $\rho\left(  \Delta S\right)  $ is a mixture unless $S$ is a
singleton event $\left\{  u_{i}\right\}  $ since $\rho\left(  \Delta\left\{
u_{i}\right\}  \right)  ^{2}=\rho\left(  \Sigma\left\{  u_{i}\right\}
\right)  ^{2}=\rho\left(  \Sigma\left\{  u_{i}\right\}  \right)  =\rho\left(
\Delta\left\{  u_{i}\right\}  \right)  $ trivially.

A \textit{partition} $\pi=\left\{  B_{1},...,B_{m}\right\}  $ on $U$ is a set
of non-empty mutually disjoint subsets $\left\{  B_{j}\right\}  _{j=1}^{m}$
whose union is $U$. \cite{ell:partitions-igpl} The partition $\pi$ is
represented by the density matrix:

\begin{center}
$\rho\left(  \pi\right)  =\sum_{j=1}^{m}\Pr\left(  B_{j}\right)  \rho\left(
\Sigma B_{j}\right)  $

Density matrix associated with a partition $\pi$ on $U$
\end{center}

\noindent that is mixed unless $\pi$ is the \textit{indiscrete partition}
$\mathbf{0}_{U}=\left\{  U\right\}  $ since $\rho\left(  \mathbf{0}%
_{U}\right)  =\rho\left(  \Sigma U\right)  $. \noindent With a suitable
interchange of rows and columns, any density matrix $\rho\left(  \pi\right)  $
defined by a partition would be block-diagonal according to the partition
blocks $B_{j}\in\pi$. For the \textit{discrete partition} $\mathbf{1}%
_{U}=\left\{  \left\{  u_{1}\right\}  ,...,\left\{  u_{n}\right\}  \right\}  $
on $U$, $\rho\left(  \mathbf{1}_{U}\right)  =\rho\left(  \Delta U\right)  $.
Thus the two extreme partitions at the top (discrete partition $\mathbf{1}%
_{U}$) and bottom (indiscrete partition $\mathbf{0}_{U}$) in the lattice of
partitions (ordered by refinement) on $U$ correspond to the two extreme
density matrices $\rho\left(  \Delta U\right)  $ and $\rho\left(  \Sigma
U\right)  $, and all the intermediate partitions $\pi$ have density matrices
that are mixtures of the pure density matrices $\rho\left(  \Sigma
B_{j}\right)  $ for their blocks. 

For the discrete partition on a subset $S$, $\mathbf{1}_{S}=\left\{  \left\{
u_{i}\right\}  \right\}  _{u_{i}\in S}$ and the indiscrete partition
$\mathbf{0}_{S}=\left\{  S\right\}  $ on a subset $S$, $\rho\left(
\mathbf{1}_{S}\right)  =\rho\left(  \Delta S\right)  $ and $\rho\left(
\mathbf{0}_{S}\right)  =\rho\left(  \Sigma S\right)  $. The discrete partition
$\mathbf{1}_{S}$ on a set $S\subseteq U$ distinguishes all the elements of $S$
from each other in singleton blocks, and thus the density matrix $\rho\left(
\mathbf{1}_{S}\right)  $ associated with that partition is the statistical
mixture of the singleton events for elements of $S$: $\rho\left(
\mathbf{1}_{S}\right)  =\rho\left(  \Delta S\right)  =\sum_{u_{i}\in S}%
\frac{p_{i}}{\Pr\left(  S\right)  }\rho\left(  \Delta\left\{  u_{i}\right\}
\right)  $. In contrast, the superposition event $\Sigma S$ associated with
$S$ represented by $\rho\left(  \Sigma S\right)  $ blurs, blobs, or coheres
together, i.e., superposes, the elements of $S$. For equal probabilities
$\frac{1}{\left\vert S\right\vert }$, the elements of $S$ are equally
superposed. Otherwise, we may say $u_{i},u_{k}\in S$ are superposed with an
\textit{amplitude} of $\rho\left(  \Sigma S\right)  _{ik}=\sqrt{\frac{p_{i}%
}{\Pr\left(  S\right)  }\frac{p_{k}}{\Pr\left(  S\right)  }}$. The entries in
the density matrices associated with $S$, namely $\rho\left(  \Delta S\right)
$ and $\rho\left(  \Sigma S\right)  $, have the same diagonal elements and
differ only in the off-diagonal elements. When an off-diagonal entry
$\rho\left(  \Sigma S\right)  _{ik}$ is non-zero, then it indicates that the
corresponding elements $u_{i},u_{k}\in S$ are cohered together with that
non-zero amplitude. All the off-diagonal elements in $\rho\left(  \Delta
S\right)  $ are zero indicating that the elements of $S$ are completely
distinguished or decohered from each other.

For a suggestive visual example, consider the outcome set $U$ as a pair of
isosceles triangles that are distinct by the labels on the equal sides and the
opposing angles.%
\begin{center}
\includegraphics[
height=1.0438in,
width=2.7319in
]%
{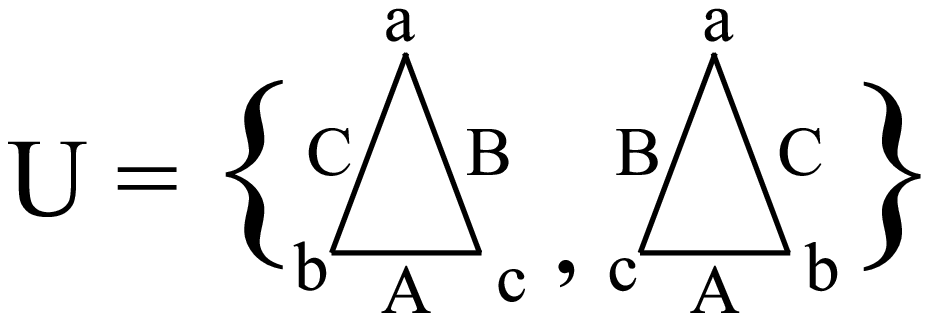}%
\\
Figure 1: Set of distinct isosceles triangles
\end{center}

\noindent The superposition event $\Sigma U$ is definite on the properties
that are common to the elements of $U$, i.e., the angle $a$ and the opposing
side $A$, but is indefinite where the two triangles are distinct, i.e., the
two equal sides and their opposing angles.%

\begin{center}
\includegraphics[
height=1.0827in,
width=3.7507in
]%
{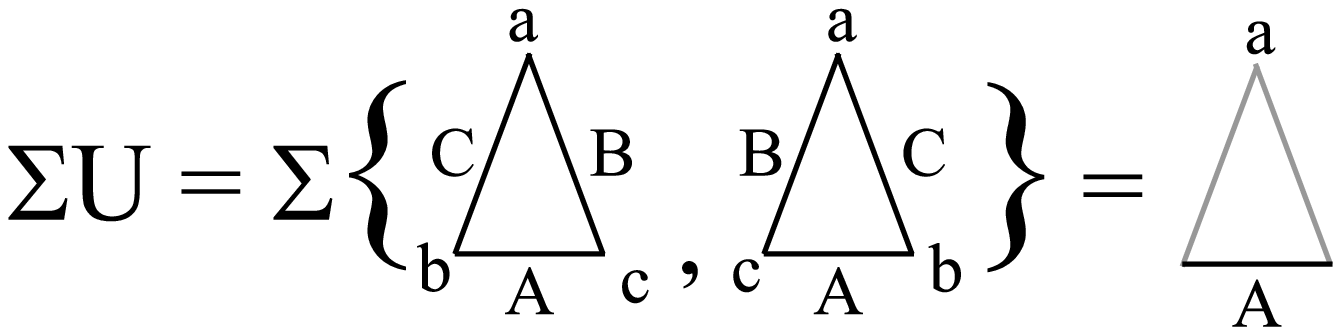}%
\\
Figure 2: The superposition event $\Sigma U$.
\end{center}

Consider the partition $\pi=\left\{  B_{1},B_{2}\right\}  =\left\{  \left\{
\diamondsuit,\heartsuit\right\}  ,\left\{  \clubsuit,\spadesuit\right\}
\right\}  $ on the outcome set $U=\left\{  \clubsuit,\diamondsuit
,\heartsuit,\spadesuit\right\}  $ with equiprobable outcomes like drawing
cards from a randomized deck. For instance, the superposition event associated
with $B_{1}=\left\{  \diamondsuit,\heartsuit\right\}  $, is pure since (rows
and columns labelled in the order $\left\{  \clubsuit,\diamondsuit
,\heartsuit,\spadesuit\right\}  $):

\begin{center}
$\rho\left(  \Sigma B_{1}\right)  =\frac{1}{\Pr\left(  \left\{  \diamondsuit
,\heartsuit\right\}  \right)  }%
\begin{bmatrix}
0 & 0 & 0 & 0\\
0 & \Pr(\left\{  \diamondsuit\right\}  ) & \sqrt{\Pr(\left\{  \diamondsuit
\right\}  )\Pr(\left\{  \heartsuit\right\}  )} & 0\\
0 & \sqrt{\Pr(\left\{  \heartsuit\right\}  )\Pr(\left\{  \diamondsuit\right\}
)} & \Pr(\left\{  \heartsuit\right\}  ) & 0\\
0 & 0 & 0 & 0
\end{bmatrix}
=%
\begin{bmatrix}
0 & 0 & 0 & 0\\
0 & \frac{1}{2} & \frac{1}{2} & 0\\
0 & \frac{1}{2} & \frac{1}{2} & 0\\
0 & 0 & 0 & 0
\end{bmatrix}
$
\end{center}

\noindent equals its square, but density matrix for the half-half mixture of
the two suit-color pure events:

\begin{center}
$\frac{1}{2}\rho\left(  \Sigma B_{1}\right)  +\frac{1}{2}\rho\left(  \Sigma
B_{2}\right)  $

$=\frac{1}{2}%
\begin{bmatrix}
0 & 0 & 0 & 0\\
0 & \frac{1}{2} & \frac{1}{2} & 0\\
0 & \frac{1}{2} & \frac{1}{2} & 0\\
0 & 0 & 0 & 0
\end{bmatrix}
+\frac{1}{2}%
\begin{bmatrix}
\frac{1}{2} & 0 & 0 & \frac{1}{2}\\
0 & 0 & 0 & 0\\
0 & 0 & 0 & 0\\
\frac{1}{2} & 0 & 0 & \frac{1}{2}%
\end{bmatrix}
=%
\begin{bmatrix}
\frac{1}{4} & 0 & 0 & \frac{1}{4}\\
0 & \frac{1}{4} & \frac{1}{4} & 0\\
0 & \frac{1}{4} & \frac{1}{4} & 0\\
\frac{1}{4} & 0 & 0 & \frac{1}{4}%
\end{bmatrix}
$
\end{center}

\noindent is a mixture since it does not equal its square.

Intuitively, the interpretation of the superposition event represented by
$\rho\left(  \Sigma B_{1}\right)  =\rho\left(  \Sigma\left\{  \diamondsuit
,\heartsuit\right\}  \right)  $ is that it is definite on the properties
common to its elements, e.g., in this case, being a red suite, but indefinite
on where the elements differ. The indefiniteness is indicated by the non-zero
off-diagonal elements that indicate that the diamond suite $\diamondsuit$ is
blurred, cohered, or superposed with the hearts suite $\heartsuit$ in the
superposition state $\Sigma\left\{  \diamondsuit,\heartsuit\right\}  $.

\section{Computing `Measurement' or Trial Probabilities with Density Matrices}

A (real-valued) \textit{random variable} on the outcome space $U$ is a
function $f:U\rightarrow%
\mathbb{R}
$ with values of $\left\{  \phi_{1},...,\phi_{m}\right\}  $. The inverse image
of $f$ is a partition $\pi=\left\{  B_{j}\right\}  _{j=1}^{m}$ where
$B_{j}=f^{-1}\left(  \phi_{j}\right)  $. In ordinary classical probability
theory, the conditional probability of getting the value $\phi_{j}$ given the
event $S$ in a trial is $\Pr\left(  \phi_{j}|S\right)  =\frac{\Pr\left(
B_{j}\cap S\right)  }{\Pr\left(  S\right)  }$. But now we have two versions of
$S$, the classical event and the superposition event. Since they have
different density matrices, we can take the given conditioning event as a
density matrix $\rho$. Let $P_{T}$ for $T\subseteq U$ be the diagonal
projection matrix with the diagonal entries $\left(  P_{T}\right)  _{ii}%
=\chi_{T}\left(  u_{i}\right)  $. Projection matrices are idempotent, i.e.,
$P_{T}P_{T}=P_{T}$ and equal their transpose $P_{T}=P_{T}^{t}$. The usual
conditional probability of the classical event $T$ given the classical event
$S$ can be computed as:

\begin{center}
$\Pr\left(  T|S\right)  :=\frac{\Pr\left(  S\cap T\right)  }{\Pr\left(
S\right)  }=\operatorname{tr}\left[  P_{T}\rho\left(  \Delta S\right)
\right]  $.
\end{center}

\noindent In general, the probability of getting the value $\phi_{j}$
conditioned by the density matrix $\rho$ is defined as:

\begin{center}
$\Pr\left(  \phi_{j}|\rho\right)  :=\operatorname{tr}\left[  P_{B_{j}}%
\rho\right]  $.
\end{center}

\noindent In particular, starting with the conditioning event being the
superposition event corresponding to $S$, that probability is:

\begin{center}
$\Pr\left(  \phi_{j}|\rho\left(  \Sigma S\right)  \right)  =\operatorname{tr}%
\left[  P_{B_{j}}\rho\left(  \Sigma S\right)  \right]  =\frac{\Pr\left(
B_{j}\cap S\right)  }{\Pr\left(  S\right)  }$.
\end{center}

\noindent This yields the perhaps surprising result that the probabilities for
the values of a random variable (or any given event $T$) are the same if the
conditioning event is the classical event $S$ represented by the mixed
$\rho\left(  \Delta S\right)  $ or the superposition event $\Sigma S$
represented by the pure $\rho\left(  \Sigma S\right)  $:

\begin{center}
$\Pr\left(  \phi_{j}|\rho\left(  \Sigma S\right)  \right)  =\operatorname{tr}%
\left[  P_{B_{j}}\rho\left(  \Sigma S\right)  \right]  =\frac{\Pr\left(
B_{j}\cap S\right)  }{\Pr\left(  S\right)  }=\operatorname{tr}\left[
P_{B_{j}}\rho\left(  \Delta S\right)  \right]  =\Pr\left(  \phi_{j}%
|\rho\left(  \Delta S\right)  \right)  $.
\end{center}

\noindent But the interpretation is quite different. The classical trial
starting with the subset $S$ represented by $\rho\left(  \Delta S\right)  $
picks out the subset $B_{j}\cap S$ represented by $\rho\left(  \Delta\left(
B_{j}\cap S\right)  \right)  $ with probability $\Pr\left(  \phi_{j}|S\right)
=\operatorname{tr}\left[  P_{B_{j}}\rho\left(  \Delta S\right)  \right]  $.
However, the `measurement' of the superposition event $\Sigma S$ represented
by $\rho\left(  \Sigma S\right)  $ `sharpens' or projects that indefinite
event to the more definite superposition event $\Sigma\left(  B_{j}\cap
S\right)  $ represented by $\rho\left(  \Sigma\left(  B_{j}\cap S\right)
\right)  $ with probability $\Pr\left(  \phi_{j}|S\right)  =\operatorname{tr}%
\left[  P_{B_{j}}\rho\left(  \Sigma S\right)  \right]  $. In either case, the
follow-up trial or `measurement' returns the same value $\phi_{j}$ with
probability $1$, i.e., $\Pr\left(  \phi_{j}|B_{j}\cap S\right)
=\operatorname{tr}\left[  P_{B_{j}}\rho\left(  \Delta\left(  B_{j}\cap
S\right)  \right)  \right]  =\operatorname{tr}\left[  P_{B_{j}}\rho\left(
\Sigma\left(  B_{j}\cap S\right)  \right)  \right]  =1$. In the classical
case, all the elements of $B_{j}\cap S$ have the value $\phi_{j}$ so the
conditioning classical event $B_{j}\cap S$ occurs with probability $1$. In the
superposition case, the property of having the value $\phi_{j}$ is definite on
the superposition event $\Sigma\left(  B_{j}\cap S\right)  $ represented by
$\rho\left(  \Sigma\left(  B_{j}\cap S\right)  \right)  $, so no `sharpening'
occurs and projection $P_{B_{j}}$ restricted to $B_{j}\cap S$ is the identity
so the measurement returns the same event $\Sigma\left(  B_{j}\cap S\right)  $
with probability $1$.

Let us illustrate this result with the case of flipping a fair coin. The
classical set of outcomes $U=\left\{  H,T\right\}  $ is represented by the
density matrix:

\begin{center}
$\rho\left(  \Delta U\right)  =$ $%
\begin{bmatrix}
\frac{1}{2} & 0\\
0 & \frac{1}{2}%
\end{bmatrix}
$.
\end{center}

%

\begin{center}
\includegraphics[
height=1.203in,
width=4.1373in
]%
{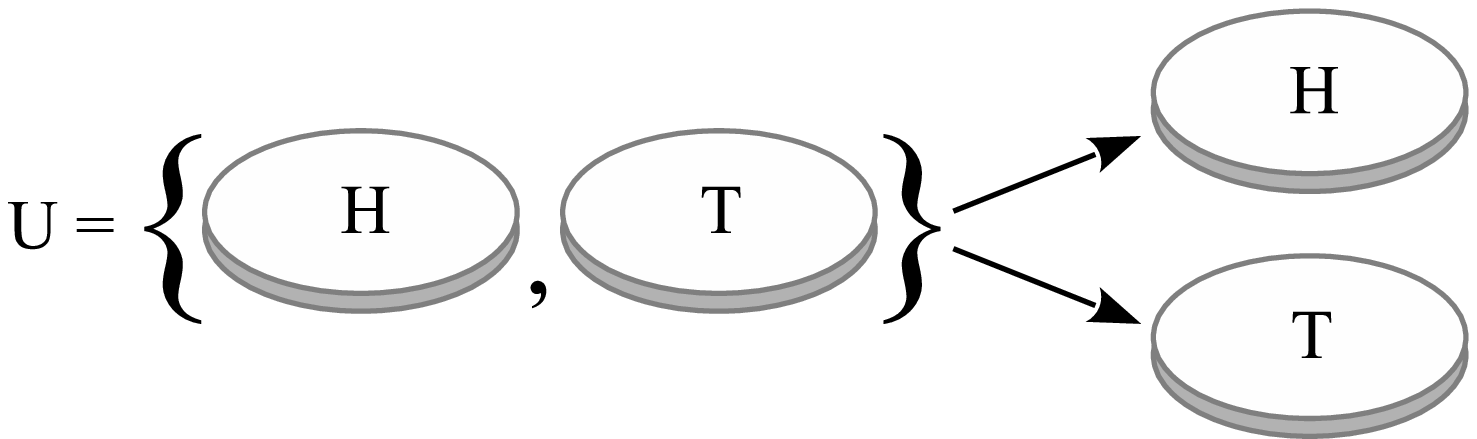}%
\\
Figure 3: Classical event: trial picks out heads or tails
\end{center}

\noindent The superposition event $\Sigma U$, that blends or superposes heads
and tails, is represented by the density matrix:

\begin{center}
$\rho\left(  \Sigma U\right)  =%
\begin{bmatrix}
\frac{1}{2} & \frac{1}{2}\\
\frac{1}{2} & \frac{1}{2}%
\end{bmatrix}
$.
\end{center}

%

\begin{center}
\includegraphics[
height=1.2834in,
width=2.7008in
]%
{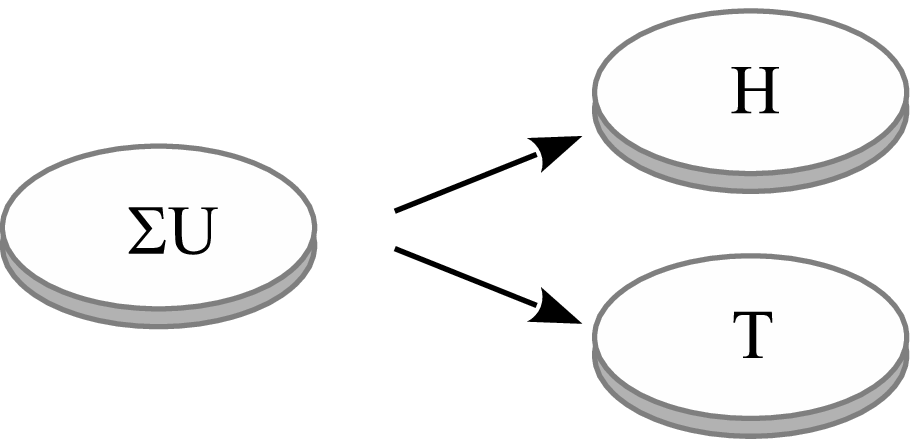}%
\end{center}

\begin{center}
Figure 4: Superposition event: Measurement sharpens to heads or tails.
\end{center}

\noindent The probability of getting heads in each case is:

\begin{center}
$\Pr\left(  H|\rho\left(  \Delta U\right)  \right)  =\operatorname{tr}\left[
P_{\left\{  H\right\}  }\rho\left(  \Delta U\right)  \right]
=\operatorname{tr}\left[
\begin{bmatrix}
1 & 0\\
0 & 0
\end{bmatrix}%
\begin{bmatrix}
\frac{1}{2} & 0\\
0 & \frac{1}{2}%
\end{bmatrix}
\right]  =\operatorname{tr}%
\begin{bmatrix}
\frac{1}{2} & 0\\
0 & 0
\end{bmatrix}
=\frac{1}{2}$

$\Pr\left(  H|\rho\left(  \Sigma U\right)  \right)  =\operatorname{tr}\left[
P_{\left\{  H\right\}  }\rho\left(  \Sigma U\right)  \right]
=\operatorname{tr}\left[
\begin{bmatrix}
1 & 0\\
0 & 0
\end{bmatrix}%
\begin{bmatrix}
\frac{1}{2} & \frac{1}{2}\\
\frac{1}{2} & \frac{1}{2}%
\end{bmatrix}
\right]  =\operatorname{tr}\allowbreak%
\begin{bmatrix}
\frac{1}{2} & \frac{1}{2}\\
0 & 0
\end{bmatrix}
=\frac{1}{2}$
\end{center}

\noindent and similarly for tails. Thus the two conditioning events $U$ and
$\Sigma U$ cannot be distinguished by performing an experiment or measurement
that distinguishes heads and tails. But this actually should not be too
surprising since the same thing occurs in quantum mechanics. For instance, a
spin measurement along, say, the $z$-axis of an electron cannot distinguish
between the superposition state $\frac{1}{\sqrt{2}}\left(  \left\vert
\uparrow\right\rangle +\left\vert \downarrow\right\rangle \right)  $ with a
density matrix like $\rho\left(  \Sigma U\right)  $ and a statistical mixture
of half electrons with spin up and half with spin down with a density matrix
like $\rho\left(  \Delta U\right)  $ \cite[p. 176]{auletta:qm}. The states can
only be distinguished by measuring in a different basis, and we will show in a
later section how probability theory with superposition events can be further
enriched to demonstrate that possibility.

It might be further noticed that the average value of a random variable can
also be computed in that same manner as in QM. If $\mathcal{O}_{f}$ is the
$n\times n$ diagonal matrix with diagonal entries $f\left(  u_{i}\right)  $
which represents $f:U\rightarrow%
\mathbb{R}
$, then the average value of the random variable restricted to a subset $S$,
$\sum_{u_{i}\in S}\Pr\left(  \phi_{i}|S\right)  f\left(  u_{i}\right)  $, is:

\begin{center}
$\left\langle f\right\rangle _{S}=\operatorname{tr}\left[  \mathcal{O}_{f}%
\rho\left(  \Delta S\right)  \right]  =\operatorname{tr}\left[  \mathcal{O}%
_{f}\rho\left(  \Sigma S\right)  \right]  $.

Average value of random variable $f$ on $S$.
\end{center}

\noindent The probability $\Pr\left(  T|S\right)  =\operatorname{tr}\left[
P_{T}\rho\left(  \Delta S\right)  \right]  =\operatorname{tr}\left[  P_{T}%
\rho\left(  \Sigma S\right)  \right]  $ is just the average value of the
characteristic function $\chi_{T}:U\rightarrow\left\{  0,1\right\}  $ on $S$
considered as a random variable on $U$, i.e., $\mathcal{O}_{\chi_{T}}=P_{T}$.
In particular,

\begin{center}
$\Pr\left(  S\right)  =\operatorname{tr}\left[  P_{S}\rho\left(  \Delta
U\right)  \right]  =\operatorname{tr}[P_{S}\rho\left(  \Sigma U\right)  ]$
\end{center}

\noindent is the average value of $\chi_{S}$ on $U$.

\section{How `Measurement' Transforms Density Matrices}

Since events, classical or superposition and any probability mixture thereof,
are now dealt with using density matrices, we need to define the resulting
change in the density matrix when a trial, an experiment, or a measurement of
a random variable occurs. Since the density matrix $\rho\left(  \Sigma
S\right)  $ is constructed as $\left\vert S\right\rangle $ times its transpose
$\left\langle S\right\vert $, the corresponding transformation by the
projection matrix $P_{T}$ is:

\begin{center}
$P_{T}\rho\left(  \Sigma S\right)  P_{T}^{t}=P_{T}\left\vert S\right\rangle
\left\langle S\right\vert P_{T}=\frac{\Pr(T\cap S)}{\Pr\left(  S\right)  }%
\rho\left(  \Sigma\left(  T\cap S\right)  \right)  $
\end{center}

\noindent since the pre- and post-multiplying by $P_{T}$ zeros all the entries
in $\left\vert S\right\rangle \left\langle S\right\vert $ except the ones
$\sqrt{\frac{p_{i}}{\Pr\left(  S\right)  }\frac{p_{k}}{\Pr\left(  S\right)  }%
}=\frac{1}{\Pr\left(  S\right)  }\sqrt{p_{i}p_{k}}$ for $u_{i},u_{k}\in T\cap
S$, and $\rho\left(  \Sigma\left(  T\cap S\right)  \right)  $ has the entries
$\frac{1}{\Pr\left(  T\cap S\right)  }\sqrt{p_{i}p_{k}}$ for the same
$u_{i},u_{k}\in T\cap S$, so $\frac{\Pr(T\cap S)}{\Pr\left(  S\right)  }%
\frac{1}{\Pr\left(  T\cap S\right)  }\sqrt{p_{i}p_{k}}=\frac{1}{\Pr\left(
S\right)  }\sqrt{p_{i}p_{k}}$ giving the result. When $T=B_{j}=f^{-1}\left(
\phi_{j}\right)  $,

\begin{center}
$P_{B_{j}}\rho\left(  \Sigma S\right)  P_{B_{j}}=\frac{\Pr(B_{j}\cap S))}%
{\Pr\left(  S\right)  }\rho\left(  \Sigma\left(  B_{j}\cap S\right)  \right)
$.
\end{center}

\noindent When the outcome of the experiment is $\phi_{j}$ with probability
$\Pr\left(  \phi_{j}|S\right)  =\frac{\Pr(B_{j}\cap S)}{\Pr\left(  S\right)
}$, then the superposition event $\Sigma S$ represented by the density matrix
$\rho\left(  \Sigma S\right)  $ is transformed into the superposition event
$\Sigma\left(  B_{j}\cap S\right)  $ represented by the density matrix
$\rho\left(  \Sigma\left(  B_{j}\cap S\right)  \right)  $. The partition
induced on $S$ by $\pi=\left\{  B_{j}\right\}  _{j=1}^{m}=\left\{
f^{-1}\left(  \phi_{j}\right)  \right\}  _{j=1}^{m}$ is $\pi\upharpoonright
S$, the partition of all the non-empty blocks $B_{j}\cap S$ for $j=1,...,m$.
The density matrix associated with all the probabilistic results is the mixed
sum of the density matrices $\rho\left(  \Sigma\left(  B_{j}\cap S\right)
\right)  $ weighted by their probabilities $\Pr\left(  \phi_{j}|S\right)
=\frac{\Pr(B_{j}\cap S)}{\Pr\left(  S\right)  }$ which is denoted by
$\rho\left(  \pi\upharpoonright S\right)  $. Thus we have:

\begin{center}
$\rho\left(  \pi\upharpoonright S\right)  :=\sum_{j=1}^{m}\Pr\left(  \phi
_{j}|S\right)  \rho\left(  \Sigma\left(  B_{j}\cap S\right)  \right)  $

$=\sum_{j=1}^{m}\frac{\Pr(B_{j}\cap S)}{\Pr\left(  S\right)  }\rho\left(
\Sigma\left(  B_{j}\cap S\right)  \right)  =\sum_{j=1}^{m}P_{B_{j}}\rho\left(
\Sigma S\right)  P_{B_{j}}$

The L\"{u}ders mixture operation: $\rho\left(  \Sigma S\right)  \leadsto
\rho\left(  \pi\upharpoonright S\right)  $.
\end{center}

The operation of experimenting with or `measuring' the random variable
$f:U\rightarrow%
\mathbb{R}
$ starting with the superposition event $\Sigma S$ represented by the pure
density matrix $\rho\left(  \Sigma S\right)  $ transforms it into the mixture
$\rho\left(  \pi\upharpoonright S\right)  =\sum_{j=1}^{m}P_{B_{j}}\rho\left(
\Sigma S\right)  P_{B_{j}}$, and that transformation is called the
\textit{L\"{u}ders mixture operation} \cite[p, 279]{auletta:qm} in quantum mechanics.

As an example, let us take $S=\left\{  \clubsuit,\diamondsuit,\spadesuit
\right\}  \subseteq U=\left\{  \clubsuit,\diamondsuit,\heartsuit
,\spadesuit\right\}  $ and take $f:U\rightarrow\left\{  0,1\right\}  \subseteq%
\mathbb{R}
$ as a random variable that distinguished the color of the suits so
$\pi=\left\{  B_{1},B_{2}\right\}  =\left\{  f^{-1}\left(  0\right)
,f^{-1}\left(  1\right)  \right\}  =\left\{  \left\{  \diamondsuit
,\heartsuit\right\}  ,\left\{  \clubsuit,\spadesuit\right\}  \right\}  $. Then
we have:

\begin{center}
$\rho\left(  \Sigma S\right)  =%
\begin{bmatrix}
\frac{1}{3} & \frac{1}{3} & 0 & \frac{1}{3}\\
\frac{1}{3} & \frac{1}{3} & 0 & \frac{1}{3}\\
0 & 0 & 0 & 0\\
\frac{1}{3} & \frac{1}{3} & 0 & \frac{1}{3}%
\end{bmatrix}
$.
\end{center}

\noindent And the probability in an experiment of getting a black suite where
$B_{2}=f^{-1}\left(  1\right)  =\left\{  \clubsuit,\spadesuit\right\}  $ is:

\begin{center}
$\Pr\left(  1|S\right)  =\operatorname{tr}\left[  B_{2}\rho\left(  \Sigma
S\right)  \right]  =\operatorname{tr}\left[
\begin{bmatrix}
1 & 0 & 0 & 0\\
0 & 0 & 0 & 0\\
0 & 0 & 0 & 0\\
0 & 0 & 0 & 1
\end{bmatrix}%
\begin{bmatrix}
\frac{1}{3} & \frac{1}{3} & 0 & \frac{1}{3}\\
\frac{1}{3} & \frac{1}{3} & 0 & \frac{1}{3}\\
0 & 0 & 0 & 0\\
\frac{1}{3} & \frac{1}{3} & 0 & \frac{1}{3}%
\end{bmatrix}
\right]  =\operatorname{tr}%
\begin{bmatrix}
\frac{1}{3} & \frac{1}{3} & 0 & \frac{1}{3}\\
0 & 0 & 0 & 0\\
0 & 0 & 0 & 0\\
\frac{1}{3} & \frac{1}{3} & 0 & \frac{1}{3}%
\end{bmatrix}
=\frac{2}{3}$.
\end{center}

The experiment of measuring the suite-colors starting with $\Sigma S$
transforms the density matrix $\rho\left(  \Sigma S\right)  $ according to the
L\"{u}ders mixture operation:

\begin{center}
$\rho\left(  \pi\upharpoonright S\right)  =\sum_{j=1}^{2}P_{B_{j}}\rho\left(
\Sigma S\right)  P_{B_{j}}=%
\begin{bmatrix}
0 & 0 & 0 & 0\\
0 & 1 & 0 & 0\\
0 & 0 & 1 & 0\\
0 & 0 & 0 & 0
\end{bmatrix}%
\begin{bmatrix}
\frac{1}{3} & \frac{1}{3} & 0 & \frac{1}{3}\\
\frac{1}{3} & \frac{1}{3} & 0 & \frac{1}{3}\\
0 & 0 & 0 & 0\\
\frac{1}{3} & \frac{1}{3} & 0 & \frac{1}{3}%
\end{bmatrix}%
\begin{bmatrix}
0 & 0 & 0 & 0\\
0 & 1 & 0 & 0\\
0 & 0 & 1 & 0\\
0 & 0 & 0 & 0
\end{bmatrix}
$

$+%
\begin{bmatrix}
1 & 0 & 0 & 0\\
0 & 0 & 0 & 0\\
0 & 0 & 0 & 0\\
0 & 0 & 0 & 1
\end{bmatrix}%
\begin{bmatrix}
\frac{1}{3} & \frac{1}{3} & 0 & \frac{1}{3}\\
\frac{1}{3} & \frac{1}{3} & 0 & \frac{1}{3}\\
0 & 0 & 0 & 0\\
\frac{1}{3} & \frac{1}{3} & 0 & \frac{1}{3}%
\end{bmatrix}%
\begin{bmatrix}
1 & 0 & 0 & 0\\
0 & 0 & 0 & 0\\
0 & 0 & 0 & 0\\
0 & 0 & 0 & 1
\end{bmatrix}
$

$=%
\begin{bmatrix}
0 & 0 & 0 & 0\\
0 & \frac{1}{3} & 0 & 0\\
0 & 0 & 0 & 0\\
0 & 0 & 0 & 0
\end{bmatrix}
+\allowbreak%
\begin{bmatrix}
\frac{1}{3} & 0 & 0 & \frac{1}{3}\\
0 & 0 & 0 & 0\\
0 & 0 & 0 & 0\\
\frac{1}{3} & 0 & 0 & \frac{1}{3}%
\end{bmatrix}
=%
\begin{bmatrix}
\frac{1}{3} & 0 & 0 & \frac{1}{3}\\
0 & \frac{1}{3} & 0 & 0\\
0 & 0 & 0 & 0\\
\frac{1}{3} & 0 & 0 & \frac{1}{3}%
\end{bmatrix}
$.
\end{center}

\section{Measurement and Logical Entropy}

The \textit{logical entropy of a partition} \cite{ellerman:newfound}
$\pi=\left\{  B_{1},...,B_{m}\right\}  $ on $U$ is:

\begin{center}
$h\left(  \pi\right)  :=\sum_{j=1}^{m}\Pr\left(  B_{j}\right)  \left(
1-\Pr\left(  B_{j}\right)  \right)  =1-\sum_{j=1}^{m}\Pr\left(  B_{j}\right)
^{2}=\sum_{j\neq j^{\prime}}\Pr\left(  B_{j}\right)  \Pr\left(  B_{j^{\prime}%
}\right)  $
\end{center}

\noindent and the logical entropy of any probability distribution $q=\left\{
q_{1},...,q_{m}\right\}  $ is similarly:

\begin{center}
$h\left(  q\right)  =1-\sum_{j=1}^{m}q_{j}^{2}=\sum_{j\neq j^{\prime}}%
q_{j}q_{j^{\prime}}=2\sum_{j<j^{\prime}}q_{j}q_{j^{\prime}}$.
\end{center}

\noindent The interpretation of the logical entropy of $\pi$ is the
probability in an ordered pair of independent draws or trials to get elements
distinguished by $\pi$ (i.e., elements from different blocks of $\pi$) or
different $q_{j}$'s. The logical entropy of any density matrix $\rho$ is:

\begin{center}
$h\left(  \rho\right)  =\operatorname{tr}\left[  \rho\left(  1-\rho\right)
\right]  =1-\operatorname{tr}\left[  \rho^{2}\right]  $.
\end{center}

\noindent The trace of any density matrix squared is the sum of all the
squared entries (or the absolute squares in complex density matrices of QM):
$\operatorname{tr}\left[  \rho^{2}\right]  =\sum_{i,k=1}^{n}\left\vert
\rho_{ik}\right\vert ^{2}$ \cite[p. 77]{fano:density}. When the partition
$\pi$ is represented by the density matrix $\rho\left(  \pi\right)  =\sum
_{j}\Pr\left(  B_{j}\right)  \rho\left(  \Sigma B_{j}\right)  $, then a simple
calculation shows that:

\begin{center}
$h\left(  \rho\left(  \pi\right)  \right)  =1-\operatorname{tr}\left[
\rho\left(  \pi\right)  ^{2}\right]  =1-\sum_{j=1}^{m}\Pr\left(  B_{j}\right)
^{2}=h\left(  \pi\right)  $.
\end{center}

Since the trace of any density matrix is $1$ and for any pure density matrix,
$\rho^{2}=\rho$, $\operatorname{tr}\left[  \rho^{2}\right]  =\operatorname{tr}%
\left[  \rho\right]  =1$ so the logical entropy of any pure density matrix is
$0$. Logical entropy measures distinctions, and in a pure superposition event
$\Sigma S$, there are no distinctions between the superposed or cohered
outcomes. When an off-diagonal element of a density matrix is non-zero, that
means the corresponding diagonal elements cohere together or are superposed in
a superposition. But when the experiment or `measurement operation'
distinguishes (or decoheres) those elements, the corresponding off-diagonal
elements are zeroed. Since the logical entropy measures distinctions, the
logical entropy created by the measurement operation can be computed as the
squares of the off-diagonal elements zeroed in the L\"{u}ders mixture
operation on the density matrices.

\begin{theorem}
The logical entropy created in the measurement of $\rho\left(  \Sigma
S\right)  $ by $\pi$, i.e. $h\left(  \rho\left(  \pi\upharpoonright S\right)
\right)  -h\left(  \rho\left(  \Sigma S\right)  \right)  $ [which equals
$h\left(  \rho\left(  \pi\upharpoonright S\right)  \right)  $ since
$\rho\left(  \Sigma S\right)  $ is pure], is the sum of the squares of the
off-diagonal elements in $\rho\left(  \Sigma S\right)  $ that are zeroed in
the L\"{u}ders mixture operation $\rho\left(  \Sigma S\right)  \leadsto
\rho\left(  \pi\upharpoonright S\right)  $.
\end{theorem}

\noindent\textbf{Proof:} \noindent All elements in the density matrix
$\rho\left(  \Sigma S\right)  $ either have the same value (e.g., all diagonal
elements and some off-diagonal elements) or are zeroed (e.g., some
off-diagonal elements) by the projections in the L\"{u}ders mixture operation.
Hence the sum of squares of the off-diagonal elements that are zeroed is:

\begin{center}
$\sum_{i,k=1}^{n}\rho\left(  \Sigma S\right)  _{ik}^{2}-\sum_{i,k=1}^{n}%
\rho\left(  \pi\upharpoonright S\right)  _{ik}^{2}=\operatorname{tr}\left[
\rho\left(  \Sigma S\right)  ^{2}\right]  -\operatorname{tr}\left[
\rho\left(  \pi\upharpoonright S\right)  ^{2}\right]  $

$=\left(  1-\operatorname{tr}\left[  \rho\left(  \pi\upharpoonright S\right)
^{2}\right]  \right)  -\left(  1-\operatorname{tr}\left[  \rho\left(  \Sigma
S\right)  ^{2}\right]  \right)  =h\left(  \pi\upharpoonright S\right)
-h\left(  \rho\left(  \Sigma S\right)  \right)  $. $\square$
\end{center}

\noindent This theorem holds, \textit{mutatis mutandis}, for quantum logical
entropy and the L\"{u}ders mixture operation in quantum information theory
where the squares are absolute squares \cite{ell-qlogicalentropy}.

To illustrate the theorem, consider the previous suite-color measurement where
$S=\left\{  \clubsuit,\diamondsuit,\spadesuit\right\}  $, The logical entropy
of the pure $\rho\left(  \Sigma S\right)  $ is $0$, and:

\begin{center}
$\rho\left(  \pi\upharpoonright S\right)  ^{2}=%
\begin{bmatrix}
\frac{1}{3} & 0 & 0 & \frac{1}{3}\\
0 & \frac{1}{3} & 0 & 0\\
0 & 0 & 0 & 0\\
\frac{1}{3} & 0 & 0 & \frac{1}{3}%
\end{bmatrix}%
\begin{bmatrix}
\frac{1}{3} & 0 & 0 & \frac{1}{3}\\
0 & \frac{1}{3} & 0 & 0\\
0 & 0 & 0 & 0\\
\frac{1}{3} & 0 & 0 & \frac{1}{3}%
\end{bmatrix}
=%
\begin{bmatrix}
\frac{2}{9} & 0 & 0 & \frac{2}{9}\\
0 & \frac{1}{9} & 0 & 0\\
0 & 0 & 0 & 0\\
\frac{2}{9} & 0 & 0 & \frac{2}{9}%
\end{bmatrix}
$
\end{center}

\noindent so $h\left(  \rho\left(  \pi\upharpoonright S\right)  \right)
=1-\operatorname{tr}\left[  \rho\left(  \pi\upharpoonright S\right)
^{2}\right]  =1-\frac{5}{9}=\frac{4}{9}$. Comparing the before and after matrices,

\begin{center}
$\rho\left(  \Sigma S\right)  =%
\begin{bmatrix}
\frac{1}{3} & \frac{1}{3} & 0 & \frac{1}{3}\\
\frac{1}{3} & \frac{1}{3} & 0 & \frac{1}{3}\\
0 & 0 & 0 & 0\\
\frac{1}{3} & \frac{1}{3} & 0 & \frac{1}{3}%
\end{bmatrix}
\leadsto%
\begin{bmatrix}
\frac{1}{3} & 0 & 0 & \frac{1}{3}\\
0 & \frac{1}{3} & 0 & 0\\
0 & 0 & 0 & 0\\
\frac{1}{3} & 0 & 0 & \frac{1}{3}%
\end{bmatrix}
=\rho\left(  \pi\upharpoonright S\right)  $,
\end{center}

\noindent we see that four entries of $\frac{1}{3}$ are zeroed (since the
different colors were distinguished by the color measurement) and the sum of
their squares is also $\frac{4}{9}$ as per the theorem. For illustrative
purposes, we might represent the matrix associated with the superposition
event $\Sigma S$ for $S=\left\{  \clubsuit,\diamondsuit,\spadesuit\right\}  $
represented by $\rho\left(  \Sigma S\right)  $ as:

\begin{center}
$%
\begin{bmatrix}
\left\{  \clubsuit,\clubsuit\right\}  & \left\{  \clubsuit,\diamondsuit
\right\}  & 0 & \left\{  \clubsuit,\spadesuit\right\} \\
\left\{  \diamondsuit,\clubsuit\right\}  & \left\{  \diamondsuit
,\diamondsuit\right\}  & 0 & \left\{  \diamondsuit,\spadesuit\right\} \\
0 & 0 & 0 & 0\\
\left\{  \spadesuit,\clubsuit\right\}  & \left\{  \spadesuit,\diamondsuit
\right\}  & 0 & \left\{  \spadesuit,\spadesuit\right\}
\end{bmatrix}
$
\end{center}

\noindent so it is clear that the four off-diagonal elements zeroed by the
measurement (that distinguished color) are the four that cohered different
colored suites together in the superposition.

The suit-color partition $\pi=\left\{  \left\{  \diamondsuit,\heartsuit
\right\}  ,\left\{  \clubsuit,\spadesuit\right\}  \right\}  $ restricted to
$S=\left\{  \clubsuit,\diamondsuit,\spadesuit\right\}  $ is $\pi
\upharpoonright S=\left\{  \left\{  \diamondsuit\right\}  ,\left\{
\clubsuit,\spadesuit\right\}  \right\}  $. In two independent ordered draws
from $S$, the probability of getting elements from different blocks of
$\pi\upharpoonright S$ is $\frac{1}{3}\frac{2}{3}+\frac{2}{3}\frac{1}{3}%
=\frac{4}{9}=h\left(  \rho\left(  \pi\upharpoonright S\right)  \right)  ,$and
that is the general interpretation of $h\left(  \pi\right)  $, the probability
in two ordered draws of getting elements in distinct blocks of $\pi$.

\section{The Pedagogical Model of Quantum Mechanics over $%
\mathbb{Z}
_{2}$}

The previous results including the fundamental theorem connecting measurement
and logical entropy hold--\textit{mutatis mutandis}--in quantum mechanics (QM)
when superposition states are being measured using a given (orthonormal) basis
$U=\left\{  u_{1},...,u_{n}\right\}  $ of an
observable.\cite{ell-qlogicalentropy} But many results in QM require
consideration of different bases. The above results about probabilities using
superposition events can be extended in the pedagogical model of quantum
mechanics over $%
\mathbb{Z}
_{2}$ (QM/Sets) \cite{ell:qm-sets} where the state space is $%
\mathbb{Z}
_{2}^{n}$ and where the $n$-ary zero-one vectors are considered as subsets of
the basis set with equiprobable outcomes. Then $U$ is just one basis which
could be taken as the computational basis, but there are many other bases. By
Gauss's formula \cite[p. 71]{lidl:finfields}, the number of ordered bases for
$%
\mathbb{Z}
_{2}^{n}$ are: $\left(  2^{n}-1\right)  \left(  2^{n}-2^{1}\right)  ...\left(
2^{n}-2^{n-1}\right)  $ and the number of unordered bases is obtained by
dividing by $n!$.

For $n=2,$ there are $\left(  2^{2}-1\right)  \left(  2^{2}-2^{1}\right)
\frac{1}{2!}=3$ (unordered) bases of $%
\mathbb{Z}
_{2}^{2}$. In the coin-flipping example where $U=\left\{  H,T\right\}  $ was
taken as the outcome set, there is another basis $U^{\prime}=\left\{
H^{\prime},T^{\prime}\right\}  $ where $\left\{  H^{\prime}\right\}  =\left\{
H,T\right\}  $ and $\left\{  T^{\prime}\right\}  =\left\{  T\right\}  $ which
is a basis since $\left\{  H^{\prime}\right\}  +\left\{  T^{\prime}\right\}
=\left\{  H,T\right\}  +\left\{  T\right\}  =\left\{  H\right\}  $ (mod $2$
addition) and $\left\{  T^{\prime}\right\}  =\left\{  T\right\}  $. The third
basis is for $U^{\prime\prime}=\left\{  H^{\prime\prime},T^{\prime\prime
}\right\}  $ where $\left\{  H^{\prime\prime}\right\}  =\left\{  H\right\}  $
and $\left\{  T^{\prime\prime}\right\}  =\left\{  H,T\right\}  $. Since we
have different bases for $%
\mathbb{Z}
_{2}^{2}$, we can consider a ket as an abstract vector that can be represented
in different bases, e.g., $\left\{  H\right\}  $, $\left\{  H^{\prime
},T^{\prime}\right\}  $, and $\left\{  H^{\prime\prime}\right\}  $ all
represent the same abstract vector in different bases. Then we can form a
ket-table where each row represents a ket. In $%
\mathbb{Z}
_{2}^{2}$, there are $2^{2}-1=3$ non-zero abstract vectors, each corresponding
to a row in the ket-table.

\begin{center}%
\begin{tabular}
[c]{|c|c|c|}\hline
$U$-basis & $U^{\prime}$-basis & $U^{\prime\prime}$-basis\\\hline\hline
$\left\{  H,T\right\}  $ & $\left\{  H^{\prime}\right\}  $ & $\left\{
T^{\prime\prime}\right\}  $\\\hline
$\left\{  H\right\}  $ & $\left\{  H^{\prime},T^{\prime}\right\}  $ &
$\left\{  H^{\prime\prime}\right\}  $\\\hline
$\left\{  T\right\}  $ & $\left\{  T^{\prime}\right\}  $ & $\left\{
H^{\prime\prime},T^{\prime\prime}\right\}  $\\\hline
\end{tabular}

Figure 5: ket-table for $%
\mathbb{Z}
_{2}^{2}$.
\end{center}

\noindent Each ket or abstract vector is a superposition in one basis and a
singleton event in the other two bases.

We saw previously that we could not distinguish the classical mixture event
$U$ associated with $\rho\left(  \Delta U\right)  $ from the superposition
event $\Sigma U$ associated with $\rho\left(  \Sigma U\right)  $ when measured
in the $U$-basis. For instance, the probability of getting heads in the two
cases is:

\begin{center}
$\Pr\left(  H|\rho\left(  \Delta U\right)  \right)  =\operatorname{tr}\left[
P_{\left\{  H\right\}  }\rho\left(  \Delta U\right)  \right]
=\operatorname{tr}\left[
\begin{bmatrix}
1 & 0\\
0 & 0
\end{bmatrix}%
\begin{bmatrix}
\frac{1}{2} & 0\\
0 & \frac{1}{2}%
\end{bmatrix}
\right]  =\operatorname{tr}%
\begin{bmatrix}
\frac{1}{2} & 0\\
0 & 0
\end{bmatrix}
=\frac{1}{2}$

$\Pr\left(  H|\rho\left(  \Sigma U\right)  \right)  =\operatorname{tr}\left[
P_{\left\{  H\right\}  }\rho\left(  \Sigma U\right)  \right]
=\operatorname{tr}\left[
\begin{bmatrix}
1 & 0\\
0 & 0
\end{bmatrix}%
\begin{bmatrix}
\frac{1}{2} & \frac{1}{2}\\
\frac{1}{2} & \frac{1}{2}%
\end{bmatrix}
\right]  =\operatorname{tr}\allowbreak%
\begin{bmatrix}
\frac{1}{2} & \frac{1}{2}\\
0 & 0
\end{bmatrix}
=\frac{1}{2}$.
\end{center}

\noindent But the two events can be distinguished when measured in a different
basis such as the $U^{\prime}$-basis.

The vector $\left\{  H\right\}  $ is expressed in the $U$-basis by the column
vector $%
\genfrac{[}{]}{0pt}{}{1}{0}%
_{U}$ (the subscript indicating the basis) and in the $U^{\prime}$-basis by
the column vector $%
\genfrac{[}{]}{0pt}{}{1}{1}%
_{U^{\prime}}$. The basis conversion matrix is

\begin{center}
$C_{U\rightarrow U^{\prime}}=%
\begin{bmatrix}
1 & 0\\
1 & 1
\end{bmatrix}
$ so $%
\begin{bmatrix}
1 & 0\\
1 & 1
\end{bmatrix}%
\begin{bmatrix}
1\\
0
\end{bmatrix}
_{U}=%
\begin{bmatrix}
1\\
1
\end{bmatrix}
_{U^{\prime}}$.
\end{center}

Hence converting the superposition $%
\genfrac{[}{]}{0pt}{}{1}{1}%
_{U}$ or $\left\{  H,T\right\}  $ to the $U^{\prime}$-basis gives:

$C_{U\rightarrow U^{\prime}}%
\begin{bmatrix}
1\\
1
\end{bmatrix}
_{U}=%
\begin{bmatrix}
1 & 0\\
1 & 1
\end{bmatrix}%
\begin{bmatrix}
1\\
1
\end{bmatrix}
_{U}=\allowbreak%
\begin{bmatrix}
1\\
0
\end{bmatrix}
_{U^{\prime}}$or $\left\{  H^{\prime}\right\}  $ so its density matrix
(computing in the reals) is $%
\begin{bmatrix}
1\\
0
\end{bmatrix}
_{U^{\prime}}\allowbreak%
\begin{bmatrix}
1 & 0
\end{bmatrix}
_{U^{\prime}}=\allowbreak%
\begin{bmatrix}
1 & 0\\
0 & 0
\end{bmatrix}
_{U^{\prime}}$. The classical mixed event $U$ is the half-half mixture of
$\left\{  H\right\}  $ and $\left\{  T\right\}  $. The basis conversion for
$\left\{  H\right\}  $ gives $C_{U\rightarrow U^{\prime}}%
\begin{bmatrix}
1\\
0
\end{bmatrix}
_{U}=%
\begin{bmatrix}
1 & 0\\
1 & 1
\end{bmatrix}%
\begin{bmatrix}
1\\
0
\end{bmatrix}
_{U}=\allowbreak%
\begin{bmatrix}
1\\
1
\end{bmatrix}
_{U^{\prime}}$ so the associated real density matrix is:

\begin{center}
$%
\begin{bmatrix}
\frac{1}{\sqrt{2}}\\
\frac{1}{\sqrt{2}}%
\end{bmatrix}
_{U^{\prime}}\allowbreak%
\begin{bmatrix}
\frac{1}{\sqrt{2}} & \frac{1}{\sqrt{2}}%
\end{bmatrix}
_{U^{\prime}}=%
\begin{bmatrix}
\frac{1}{2} & \frac{1}{2}\\
\frac{1}{2} & \frac{1}{2}%
\end{bmatrix}
_{U^{\prime}}$
\end{center}

\noindent and for $\left\{  T\right\}  $, $C_{U\rightarrow U^{\prime}}%
\begin{bmatrix}
0\\
1
\end{bmatrix}
_{U}=%
\begin{bmatrix}
1 & 0\\
1 & 1
\end{bmatrix}%
\begin{bmatrix}
0\\
1
\end{bmatrix}
_{U}=\allowbreak%
\begin{bmatrix}
0\\
1
\end{bmatrix}
_{U^{\prime}}$ so its real density matrix is:

\begin{center}
$%
\begin{bmatrix}
0\\
1
\end{bmatrix}
_{U^{\prime}}\allowbreak%
\begin{bmatrix}
0 & 1
\end{bmatrix}
_{U^{\prime}}=\allowbreak%
\begin{bmatrix}
0 & 0\\
0 & 1
\end{bmatrix}
_{U^{\prime}}$.
\end{center}

\noindent Their half-half mixture has the density matrix in the $U^{\prime}$-basis:

\begin{center}
$\frac{1}{2}%
\begin{bmatrix}
\frac{1}{2} & \frac{1}{2}\\
\frac{1}{2} & \frac{1}{2}%
\end{bmatrix}
_{U^{\prime}}+\frac{1}{2}%
\begin{bmatrix}
0 & 0\\
0 & 1
\end{bmatrix}
_{U^{\prime}}=%
\begin{bmatrix}
\frac{1}{4} & \frac{1}{4}\\
\frac{1}{4} & \frac{3}{4}%
\end{bmatrix}
_{U^{\prime}}$.
\end{center}

We then measure by the partition $\sigma=\left\{  \left\{  H^{\prime}\right\}
,\left\{  T^{\prime}\right\}  \right\}  $ with half-half probabilities so the
probability of $H^{\prime}$ for the superposition event $\left\{  H,T\right\}
$ or $\left\{  H^{\prime}\right\}  $ in the $U^{\prime}$-basis is:

\begin{center}
$\operatorname{tr}\left[  P_{\left\{  H^{\prime}\right\}  }%
\begin{bmatrix}
1 & 0\\
0 & 0
\end{bmatrix}
_{U^{\prime}}\right]  =\operatorname{tr}\left[
\begin{bmatrix}
1 & 0\\
0 & 0
\end{bmatrix}
_{U^{\prime}}%
\begin{bmatrix}
1 & 0\\
0 & 0
\end{bmatrix}
_{U^{\prime}}\right]  =\operatorname{tr}%
\begin{bmatrix}
1 & 0\\
0 & 0
\end{bmatrix}
_{U^{\prime}}=1$
\end{center}

\noindent and for the classical mixture of half $\left\{  H\right\}  $ and
half $\left\{  T\right\}  $which in the $U^{\prime}$-basis is the mixture of
half $\left\{  H^{\prime},T^{\prime}\right\}  $and half $\left\{  T^{\prime
}\right\}  $, is:

\begin{center}
$\operatorname{tr}\left[  P_{\left\{  H^{\prime}\right\}  }%
\begin{bmatrix}
\frac{1}{4} & \frac{1}{4}\\
\frac{1}{4} & \frac{3}{4}%
\end{bmatrix}
_{U^{\prime}}\right]  =\operatorname{tr}\left[
\begin{bmatrix}
1 & 0\\
0 & 0
\end{bmatrix}
_{U^{\prime}}%
\begin{bmatrix}
\frac{1}{4} & \frac{1}{4}\\
\frac{1}{4} & \frac{3}{4}%
\end{bmatrix}
_{U^{\prime}}\right]  =\operatorname{tr}%
\begin{bmatrix}
\frac{1}{4} & \frac{1}{4}\\
0 & 0
\end{bmatrix}
_{U^{\prime}}=\frac{1}{4}$.
\end{center}

\noindent The first calculation makes intuitive sense since the superposition
$\left\{  H,T\right\}  $ in the $U$-basis is the singleton event $\left\{
H^{\prime}\right\}  $ in the $U^{\prime}$-basis, so measuring in the
$U^{\prime}$-basis for the event $\left\{  H^{\prime}\right\}  $ will give
$\left\{  H^{\prime}\right\}  $ with probability $1$. The second calculation
makes intuitive sense since it is half-half in the mixture whether we get the
$\left\{  T^{\prime}\right\}  $ event or the $\left\{  H^{\prime},T^{\prime
}\right\}  $ event and then the probability of getting $H^{\prime}$ is zero
for the $\left\{  T^{\prime}\right\}  $ event and $\frac{1}{2}$ for the
$\left\{  H^{\prime},T^{\prime}\right\}  $ event so the overall probability of
$\left\{  H^{\prime}\right\}  $ is $\frac{1}{4}$. Thus the two events, the
classical mixture of half $\left\{  H\right\}  $ and half $\left\{  T\right\}
$, and the superposition $\left\{  H,T\right\}  $, which cannot be
distinguished by measurements in the $U$-basis, can be distinguished by
measurement in the $U^{\prime}$-basis.

\section{Concluding Remarks}

Ordinary finite probability theory can be extended to include superposition
events by using the two-dimensional representations of:

\begin{itemize}
\item $\rho\left(  \Delta S\right)  $ for the classical event $S\subseteq U$,
where the outcomes in $S$ are kept discrete and completely decohered, and

\item $\rho\left(  \Sigma S\right)  $ for the superposition event $\Sigma S$
that superposes or coheres together the outcomes in $S$.
\end{itemize}

The calculation of probabilities for classical events in ordinary finite
probability theory can be computed using the density matrices in the form
$\rho\left(  \Delta S\right)  $ for classical events $S$. Thus the extension
to include superposition events just extends to using density matrices of the
form $\rho\left(  \Sigma S\right)  $, and the density matrix formalism also
represents classical mixtures of superposition events.

Ordinary finite probability theory sticks with one outcome or sample space
$U$. But the whole machinery can be developed in $%
\mathbb{Z}
_{2}^{n}$ where $U$ is just one among many basis sets and then it is part of
the pedagogical model of quantum mechanics over $%
\mathbb{Z}
_{2}$ or QM/Sets. That pedagogical model of QM over $%
\mathbb{Z}
_{2}$ could also be viewed as just the \textit{non-commutative extension of
finite probability theory with superposition events} (since the bases do not
in general commute in QM/Sets). Many characteristic QM results can be modeled
in this non-commutative probability theory such as the double-slit experiment,
the indeterminacy principle, quantum statistics for identical particles, and
even Bell's Theorem.\cite{ell:qm-sets}

Our purpose has been to illustrate, in a rather classical setting, the notion
of a superposition event, where all the outcomes in the event cohere together
(with various amplitudes), so the event is objectively indefinite between
those outcomes. The notions of objective-indefiniteness and superposition are
the essentials in what Abner Shimony called the "Literal" or
objectively-indefinite interpretation of QM, an interpretation that is
routinely neglected in the literature that focuses on fantasies about many
worlds or hidden variables.

\begin{quotation}
\noindent From these two basic ideas alone -- indefiniteness and the
superposition principle -- it should be clear already that quantum mechanics
conflicts sharply with common sense. If the quantum state of a system is a
complete description of the system, then a quantity that has an indefinite
value in that quantum state is objectively indefinite; its value is not merely
unknown by the scientist who seeks to describe the system. \cite[p.
47]{shim:reality}

\noindent But the mathematical formalism ... suggests a philosophical
interpretation of quantum mechanics which I shall call "the Literal
Interpretation." ...This is the interpretation resulting from taking the
formalism of quantum mechanics literally, as giving a representation of
physical properties themselves, rather than of human knowledge of them, and by
taking this representation to be complete. \cite[pp. 6-7]{shim:vienna}
\end{quotation}

\noindent To understand or interpret QM, one needs to better understand the
notions of objective indefiniteness and superposition as well as the related
notion of a (distinguishing) measurement that sharpens an indefinite
superposition event to a mixture of more definite ones. We have shown that the
concepts of superposition, objective-indefiniteness, and measurement can be
illustrated in a very small extension of classical finite probability
theory--which should help to intuitively understand those notions in quantum mechanics.

\end{document}